# Citing and Reading Behaviours in High-Energy Physics.
## How a Community Stopped Worrying about Journals and Learned to Love Repositories

Anne Gentil-Beccot[1], Salvatore Mele[2]
CERN, European Organization for Nuclear Research
CH1211, Genève 23, Switzerland

Travis C. Brooks[3]
SLAC National Accelerator Laboratory[4]
Sand Hill Road, Menlo Park, CA 94309, United States of America

## Abstract

Contemporary scholarly discourse follows many alternative routes in addition to the three-century old tradition of publication in peer-reviewed journals. The field of High-Energy Physics (HEP) has explored alternative communication strategies for decades, initially via the mass mailing of paper copies of preliminary manuscripts, then via the inception of the first online repositories and digital libraries.

This field is uniquely placed to answer recurrent questions raised by the current trends in scholarly communication: is there an advantage for scientists to make their work available through repositories, often in preliminary form? Is there an advantage to publishing in Open Access journals? Do scientists still read journals or do they use digital repositories?

The analysis of citation data demonstrates that free and immediate online dissemination of preprints creates an immense citation advantage in HEP, whereas publication in Open Access journals presents no discernible advantage. In addition, the analysis of clickstreams in the leading digital library of the field shows that HEP scientists seldom read journals, preferring preprints instead.

## 1. Introduction

The last two decades have heralded major changes in scholarly communication. Electronic journals have increased the visibility and accessibility of scientific information. At the same time, free online resources have enabled the dissemination of some versions of scholarly articles either before publication, or simultaneously, or after some embargo, further facilitating scientific dialogue. The interaction between these dissemination models spans the spectrum from synergy to hostility, and has generated debates ranging from the economics of scientific publishing to the sociology of scientific discourse. All stakeholders of the scholarly communication process are involved in this debate: from librarians who need parameters on which to base collection development strategies to policy makers who weigh the benefits of Open Access, and from scientific publishers developing business models and platforms to scientists as users of scientific information.

[1] Anne.Gentil-Beccot@cern.ch
[2] Salvatore.Mele@cern.ch
[3] travis@slac.stanford.edu
[4] Work partly supported by Department of Energy contract DE--AC02--76SF00515

Fact-based evidence is sometimes scant in this debate, and this want has heightened interest in understanding the reading and citing behaviours of scholars. The field of High-Energy Physics (HEP) offers a unique environment in which to study these habits due to several factors. First and foremost, the Open Access culture of the field dates back decades, to when scholars sent preprints (manuscripts of their publications which had not yet appeared in peer-reviewed journals) to their peers around the world [1,2,3]. Research libraries at the large laboratories in the field indexed and classified these resources, which gave rise to SPIRES, to our knowledge the first electronic catalogue of grey literature. SPIRES not only collected all preprints in the field, but was also updated with information upon publication of a preprint [4,5]. Incidentally, SPIRES was also the first web server in the U.S. and the first database on the web [6]. In 1991, Paul Ginsparg, then at the Los Alamos National Laboratory in New Mexico, conceived arXiv, an internet-based system to disseminate preprints [7]. arXiv was first based on e-mail and then on the web, becoming the first repository and the first "green" Open Access[5] platform.

Today the coalition of SPIRES and arXiv seamlessly serves the information needs of the HEP community [8]. A study of these resources sheds light on three main questions: is there an advantage for scientists to make their work available through repositories, often in preliminary form? Is there an advantage to publish in Open Access journals? Do scientists still read journals or do they use digital repositories?

This article is structured as follows. Section 2 gives further background on scholarly communication in HEP and presents the data sets used. The results of an analysis of SPIRES data on the citation behaviour of HEP scientists is presented in Section 3 and Section 4, demonstrating the "green" Open Access advantage in HEP. The possible existence of a "gold" Open Access[6] advantage is addressed in section 5. Finally, Section 6 presents a direct analysis of the clickstreams of SPIRES users, which sheds light on the reading habits of HEP scientists. Section 7 summarizes the findings of these analyses in the wider framework of the way scientific discourse has evolved in HEP.

---

[5] With the term "green" Open Access we denote the free online availability of scholarly publications in a repository. In the case of HEP, the submission to these repositories, typically arXiv, is not mandated by universities or funding agencies, but is a free choice of authors seeking peer recognition and visibility.
[6] With the term "gold" Open Access we denote the free online availability of a scholarly publication on the web site of a scientific journals.

## 2. Methodology and Background

All data used in this article are extracted from the SPIRES database. SPIRES is a database of metadata that has covered all of the HEP literature, in both published and preprint form, since 1974[7]. It contains over 750,000 records. It is hosted at the SLAC National Accelerator Laboratory in California, and compiled jointly with DESY, the Deutsches Elektronen-Synchrotron in Hamburg, Germany, and Fermilab, the Fermi National Accelerator Laboratory in Illinois [4,5,9]. SPIRES compiles metadata for the entire corpus of HEP literature from sources such as arXiv, peer-reviewed journals, conference websites and selected institutional repositories. In particular, it maintains citation data, keywords, classifications and authors with their institutional affiliations for HEP articles that appeared on arXiv or in journals.

The HEP community relies heavily on communication through preprints, and therefore SPIRES counts citations to and from preprints. In SPIRES, citations to preprints are aggregated with the citations to the published versions, once available, treating the two versions as a single entity. This differs from most bibliometric approaches which only consider citations from published articles to published articles, and this feature is crucial for the conclusions of this study. It is also worth remarking that SPIRES considers only content relevant to HEP. This creates a closed ecosystem in which to analyse the citing behaviour of the HEP community, in a noiseless and controlled environment.

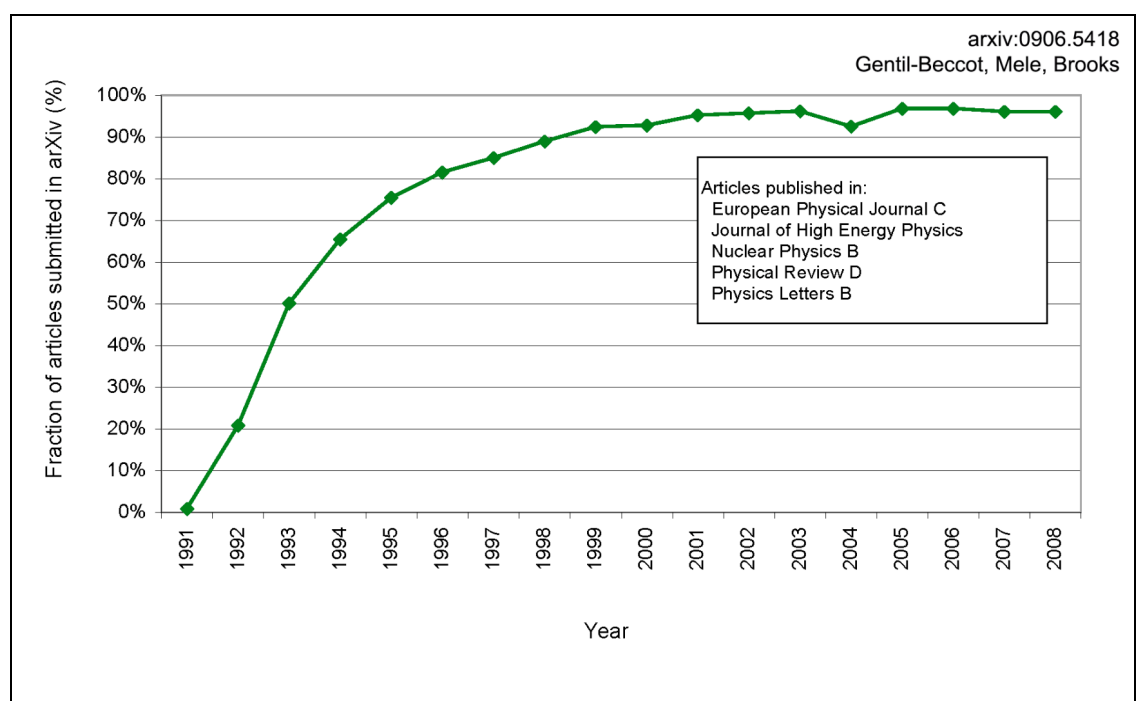

*Figure 1. Fraction of articles published in the main peer-reviewed HEP journals which also appeared, in some version, on arXiv.org as a function of time.*

---

[7] SPIRES is about to be replaced by a new platform, INSPIRE, jointly realised by the SPIRES partners (DESY, Fermilab, and SLAC) and CERN, the European Organization for Nuclear Research. INSPIRE will add novel functionalities to SPIRES, such as search speed, full-text search, text- and data-mining capabilities, and capture of user-generated content.
More details are available at www.projecthepinspire.net.

The arXiv.org [10] repository also plays a crucial role in HEP, storing the vast majority of preprints of the field. Figure 1 presents the time evolution of the fraction of the content of the main peer-reviewed HEP journals[8] that is also available in a preprint form on arXiv. For over a decade, between 90% and 100% of published articles have been provided by arXiv[9]. It is worth noting that many HEP scientists routinely upload to arXiv a revised version of their preprint which matches the final peer-reviewed version, including any corrections introduced during the publication process.

The combination of SPIRES and arXiv provides complete coverage of the HEP literature, SPIRES providing detailed metadata, and arXiv providing full-text preprint versions of nearly all journal articles. A comprehensive survey of HEP practitioners established that nearly 90% of them rely primarily on SPIRES and arXiv as their point of entry to the literature [8].

## 3. Is there an advantage for scientists to make their work available through repositories, often in preliminary form?

A potential citation advantage due to the appearance of a scientific article in a repository has been discussed within the wider Open Access debate [13,14,15]. The study of HEP scientists, with their comprehensive adoption of arXiv as a subject repository, allows the determination of the advantages that come with this choice, or, in turn, which incentives drove such a wide adoption of arXiv. HEP journal articles and preprints, excluding conference proceedings, which appeared from 1991 to 2007 and corresponding to 286'180 manuscripts, are split into three mutually exclusive sets.
1. Articles which were only submitted to arXiv and never published.
2. Articles which were published without appearing on arXiv.
3. Articles which were published and also appeared on arXiv.

The population of these three sets of articles changes dramatically with time. Set 1 varies from 0.6% of the total in 1991, to 34.8% of the total in 2008. Set 2 from 95.8% to 12.8% and set 3 from 3.6% to 52.4%. The impact of articles in each of these sets was parametrized by means of the Impact Factor (IF) [16]. The IF is computed using SPIRES data to calculate the number of citations collectively received during year $n$, by the articles of each set which appeared in years $n$-1 and $n$-2. For articles in set 1, the date of appearance corresponds to the date of appearance on arXiv. For articles in set 2, it corresponds to the date of publication. For articles in set 3 it corresponds to the earliest date, either that of submission to arXiv or that of publication.

Figure 2 presents the IF for the three sets of articles as a function of the year in which citations are counted. A marked citation advantage is present for set 3. For 2008, this advantage is a factor five: the IF for articles both published and also available on arXiv is collectively five times larger than for articles in sets 1 and 2, which are just submitted to arXiv or just published, respectively.

---

[8] In HEP, five journals publish the vast majority of the articles of the field: *European Physical Journal C*, *Journal of High Energy Physics*, *Nuclear Physics B*, *Physics Letters B*, *Physical Review D*. They contain mostly HEP content. Another journal, *Physical Review Letters*, only carries about 10% of HEP content. It is not included in Figure 1, however its content is almost entirely on arXiv. [11, 12]

[9] The slight decrease in the year 2004 is due to a large conference proceedings published in one journal (*European Physical Journal C*) whereby most of this material was not submitted to arXiv, as sometimes customary in HEP for some conference proceedings built on existing material.

The low IF for recent material that is only published could potentially be due to a selection effect wherein authors who choose not to use arXiv, by 2000 the mainstream mode of communication for HEP, are also less likely to produce highly cited papers. It is also of similar interest to compare the situation with the early years of arXiv. In 1993, the first IF data point after the inception of arXiv in 1991, a relatively small fraction of articles were submitted to arXiv. Notwithstanding the relative novelty of arXiv at that time, articles that were both submitted to arXiv and then published, already had a citation advantage of a factor two! Several explanations for this observation are possible: a genuine advantage deriving from larger dissemination; influential, and therefore highly-cited, early adopters; the preference to publicly expose on arXiv only the work that authors felt was of a higher quality. The IF of manuscripts only submitted to arXiv was very low in the early years, possibly reflecting a reluctance to cite material prior to peer-review. A more detailed analysis would be needed to ascertain these historical and sociological effects, which transcend the scope of this article.

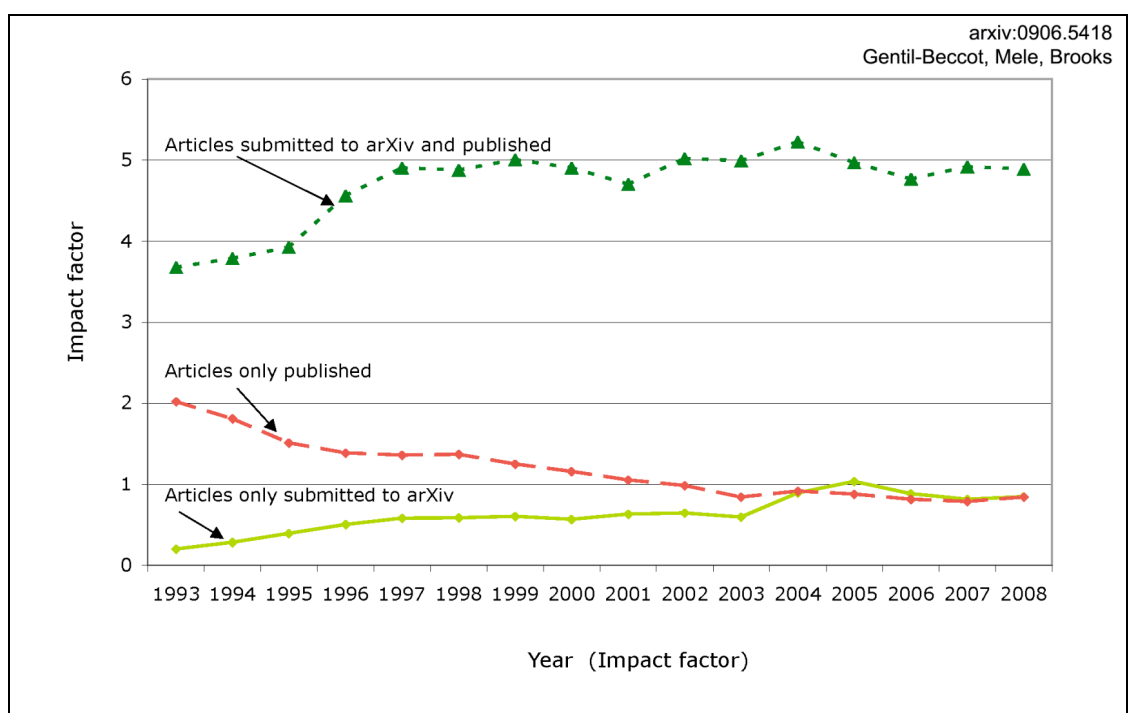

*Figure 2. Evolution of the Impact Factor of three sets of HEP articles as a function of the year for which the Impact Factor is calculated. Those which were only submitted to arXiv and never published, those which were published without having ever been submitted to arXiv and those which were published and also appeared on arXiv.*

## 4. The real advantage: immediacy

The results in Section 3 demonstrate an immense citation advantage for articles submitted to arXiv. To understand the incentives for HEP scientists to use arXiv, and to contribute to the debate on embargoed Open Access, it is interesting to investigate the origin of this advantage. Is it due to a wider or an earlier dissemination? To answer this question 26,741 articles published in two leading HEP journals are considered. These appeared from 1998 to 2007 in the *Journal of High Energy Physics* and *Physical*

*Review D*[10] and are split in two samples, those which were submitted to arXiv (96.4% of the total) and those which were published without appearing on arXiv (3.6% of the total).

Figure 3 presents the average number of citations for articles in the two sets as a function of the time of the citation, relative to the time of publication. Articles submitted to arXiv begin accumulating citations at negative times (prior to publication). The sample of articles which were submitted to arXiv is much larger, and thus it has smaller variation and smoother behaviour, except for articles with a large time-gap before publication, which are rare.

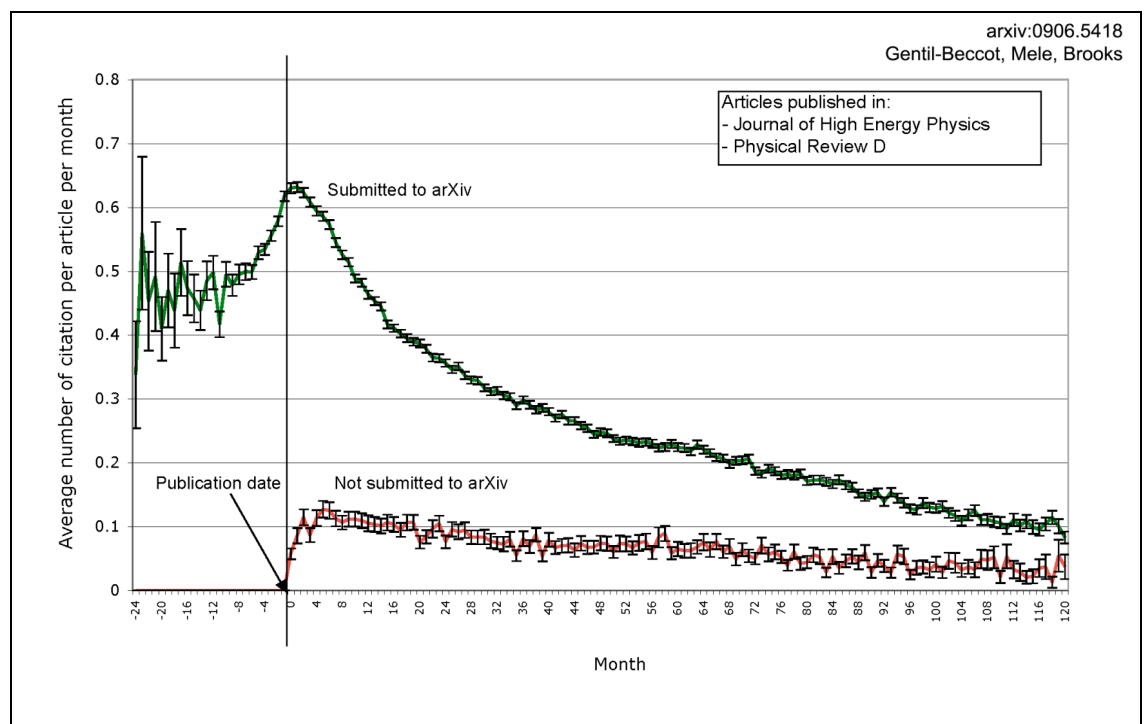

*Figure 3. Average number of citations per article per month as a function of the time of the citation relative to the time of publication. Citations at negative times occurred while articles were in their preprint form. Citations at positive times occurred after the publication of the articles. Data is from 26741 articles from the Journal of High Energy Physics and Physical Review D over the period from 1998 to 2007. The vertical bars represent the statistical uncertainty of the data, and are calculated as $1/N (\Sigma_i(x_i-<x>)^2)^{1/2}$, where the sum runs over the N articles receiving $x_i$ citations in the same month after publication and $<x>$ is the mean number of citations for articles at this time after publication.*

Articles that were submitted to arXiv prior to publication show an immense Open Access advantage, which appears here as a much larger area under their citation curve. This is of course consistent with the fact that in HEP the scientific discourse happens on arXiv. In addition, Figure 3 demonstrates the time advantage of articles submitted to arXiv. Citation begins well before publication occurs, as seen from the large amounts of citations to arXiv papers at negative times. The difference in the shape of the two curves also signifies that many citations received in the first few months after publication occur because authors read the preprint earlier.

---

[10] We restrict our analysis to just 2 of the 5 larger journals of the field in order to simplify data handling. However, it is worth remarking that these two journals collectively cover about ½ of the published HEP literature and are therefore representative of the behaviour of the entire data set.

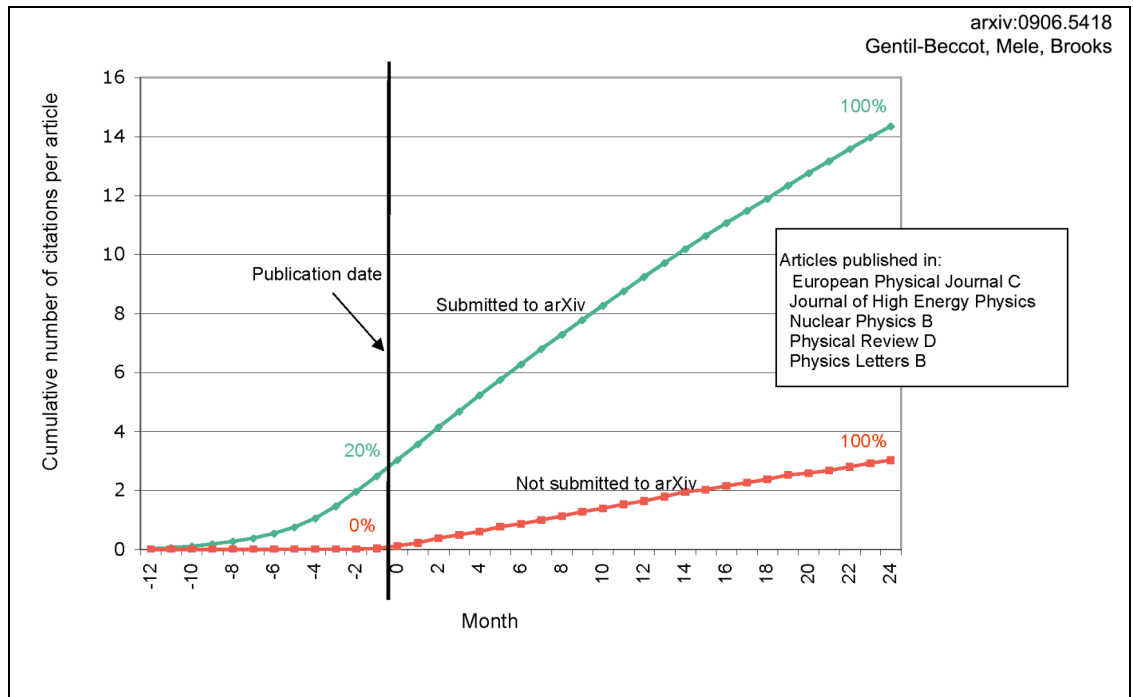

*Figure 4. Cumulative citation count as a function of the age of the paper relative to its publication date. 4839 articles from 5 major HEP journals published in 2005 are considered.*

Figure 4 presents the cumulative number of citations per article as a function of citation time relative to the publication time. The 4839 articles which appeared in the five major HEP journals in 2005 are considered. They are separated into two groups, those that were submitted to arXiv and those that were not. At the time of publication, articles submitted to arXiv have already attained 20% of the total number of citations they will eventually collect by the end of the two following years. Obviously, articles not submitted to arXiv have no citation at the time of publication. The arXiv preprints, when published, have already amassed an advantage that non-arXiv articles can never recoup.

These findings demonstrate that HEP scientists do not wait for an article to be published before citing it. The use of paper preprints for decades, and arXiv since 1991, enables scientific discussion in HEP to begin as soon as possible, accelerating the scientific process as well as the communication process.

## 5. Is there an advantage to publishing in Open Access journals?

Some studies suggest that "gold" Open Access articles have a citation advantage [15, 17], *i.e.* that articles published in journals which are freely accessible are cited more frequently than articles appearing in journals whose access is behind subscription barriers. The underlying hypothesis is that a freely-available article will be read more, and thus cited more, than an article which is not freely available. Other studies report no evidence of this effect [18,19,20]. Which is the situation in HEP, where an immense

Open Access advantage is already provided by arXiv? To answer this question we investigate a special set of Open Access articles, and a corresponding control sample.

The *Journal of High Energy Physics* (JHEP), currently jointly published by SISSA and IOPP, carries about 20% of the HEP literature [11,12]. As of 2007, JHEP has offered an Open Access option to some subscribing institutions, allowing articles by all authors affiliated with these institutions, to be published as Open Access on the publisher's site [21]. In 2007 and 2008, this option has applied to all articles with at least one author from CERN, DESY, Fermilab, SLAC or any French HEP institute. This sample amounts to about 25% of the content of JHEP.

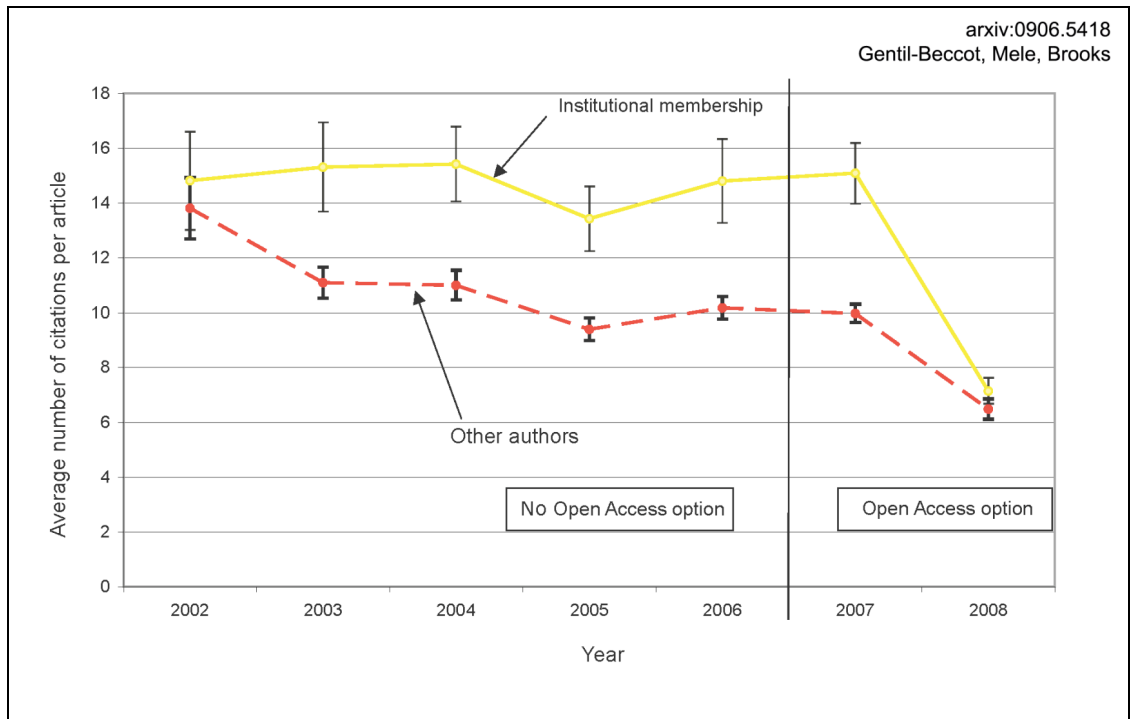

*Figure 5: Average number of citations per article in JHEP as a function of time. From 2007 onwards, articles are split in Open Access and non Open Access. Prior to 2007, articles are split according to whether their authors are affiliated with institutions that offered Open Access as of 2007.*

The right panel of Figure 5 compares the average citation count for two different samples of articles published in JHEP between January 2007 and December 2008: the Open Access fraction and the fraction which is not Open Access. While articles published in 2008 had little time to cumulate citations at the time of the analysis, the citation count in 2007 seems instead to suggest an Open Access advantage.

This effect is further investigated by considering articles published in JHEP between 2002 and 2006, before the launch of this Open Access policy. This sample is also divided in two parts. Articles authored by at least one author affiliated to one of the institutions that were to offer Open Access as from 2007, and all remaining articles. The left panel of Figure 5 presents the average citation count for these two additional samples[11]. This data sample shows the same advantage observed as for the period 2007-2008. Although the data cover a short time span and have limited statistics, they suggest no discernible increase in citation advantage for articles from the selected institutions after they began providing Open Access. The advantage in 2007-2008 could easily be attributed to the fact that papers from these institutions get more citations in general.

In conclusion, we do not detect any citation advantage from publication in Open Access journals in HEP. This finding is similar to the results obtained in the fields of Astrophysics [22], Condensed Matter [23], Mathematics [24].

## 6. Do scientists still read journals?

The citation analysis presented in the previous sections gives an understanding of the speed and manner of the scientific discourse in HEP, but is still removed from the question of what scientists actually read. Do HEP scientists still read journals? Recent studies [8] found that about 50% of HEP scientists turn to SPIRES for a bibliographic search. Therefore, the analysis of clickstreams of SPIRES users once an article has been identified and can be accessed, gives a clear representation of the reading habits of the community. A bibliographic search in SPIRES results in a list of all available links where the articles can be retrieved, most likely on arXiv and publisher website, as applicable.

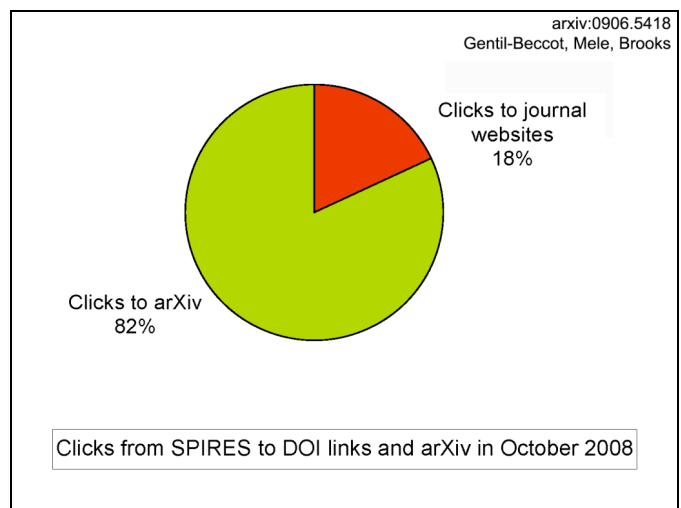

*Figure 6. Relative frequency of outgoing clicks from SPIRES if a record returned by a bibliographic search links to both a preprint on arXiv and a journal article hosted on a publisher web site.*

SPIRES clickstreams collected during October 2008 are analysed. The study is restricted to clicks that occurred from records displaying both a link to arXiv and to a publisher website. Figure 6 compares the frequency of a click on a publisher website with a click to arXiv. arXiv is preferred by more than a factor four. This result is found to be mostly independent of the publisher, the journal and the age of the article.

---

[11] The analysis is restricted to citations appearing in the first two years after publication, since JHEP has an Open Access embargo policy, which makes all articles Open Access two years after they are published.

The reasons for preferring the preprint over the journal published version, when there is such a choice, are not immediately apparent. Several characteristics of the field are relevant to explain this phenomenon. First and foremost, in HEP it is standard practice to (re)submit author-formatted versions of an article to arXiv upon acceptance to a journal, so that often arXiv presents a version very similar, or entirely equivalent, to the published one. Additionally, since preprints are available via arXiv long before the journal version, HEP physicists might have an ingrained habit to turn to arXiv for their research. It is also worth noting that the links to publishers' web pages are usually directed to a "splash page" that contains an abstract or other similar information, from which an additional click is needed to access the full text of the article. At the same time, arXiv links from SPIRES take the user directly to the full text of the article: this shorter path might influence the user decision.

Finally, the arXiv version is freely available, while the journal version is most likely accessible only to journal subscribers. Most users of SPIRES are scientists that usually have access to HEP journals, but might not enjoy this access outside their workplace.

There are as many physicists using arXiv directly for their bibliographic searches as those who use SPIRES, while the fraction that uses publishers' web sites directly is negligible [8]. Taking this additional cohort into account, the advantage of arXiv over the published version, quoted above as a factor four, is therefore conservative and might be closer to a factor eight.

## 7. Conclusions

Scholarly communication is at a cross road of new technologies and publishing models. The analysis of almost two decades of use of preprints and repositories in the HEP community provides unique evidence to inform the Open Access debate, through four main findings:
1. Articles submitted to an Open Access subject repository, arXiv, receive 5 times more citations than articles which are not.
2. The citation advantage of articles appearing in a repository is connected to their dissemination prior to publication, 20% of citations of HEP articles over a two-year period occur before publication.
3. No discernable citation advantage can yet be observed in the statistically-limited sample of articles published in "gold" Open Access journals.
4. HEP scientists are between four and eight times more likely to download an article in its preprint form from arXiv rather than its final published version on a journal web site.

Taken together these findings lead to three general conclusions about scholarly communication in HEP, as a discipline that has long embraced green Open Access:

1. There is an immense advantage for individual authors, and for the discipline as a whole, in free and immediate circulation of ideas, resulting in a faster scientific discourse.
2. The advantages of Open Access in HEP come without mandates and without debates. Universal adoption of Open Access follows from the immediate benefits for authors.
3. Peer-reviewed journals have lost their role as a means of scientific discourse, which has effectively moved to the discipline repository.

HEP has charted the way for a possible future in scholarly communication to the full benefit of scientists, away from over three centuries of tradition centred on scientific journals. However, HEP peer-reviewed journals play an indispensable role, providing independent accreditation, which is necessary in this field as in the entire, global, academic community. The next challenge for scholarly communication in HEP, and for other disciplines embracing Open Access, will be to address this novel conundrum. Efforts in this direction have already started, with initiatives such as SCOAP$^3$ [12,25].

## Acknowledgements

We are indebted to Carmen van Pamel and Nicholas Steketee, on an internship from the Collège du Léman, for their collaboration in the analysis of the data published in this article.

[24] P.M. DAVIS, M.J. FROMERTH, Does the arXiv lead to higher citations and reduced publisher downloads for mathematics articles?, Scientometrics, 71 (2007) 202 – 215, arXiv:cs/0603056

[25] The SCOAP3 project is described at: http://scoap3.org [Last visited June 28 2009]